\documentclass[%
 reprint,
 amsmath,amssymb,
 aps,
prb,
floatfix,
showkeys
]{revtex4-2}
\usepackage{xcolor}
\usepackage{graphicx}
\usepackage{dcolumn}
\usepackage{bm}
\usepackage{hyperref}
\usepackage{ulem}

\hypersetup{
  colorlinks   = true, 
  urlcolor     = green, 
  linkcolor    = blue, 
  citecolor   = blue 
}

\begin{document}

\preprint{?}

\title{Dynamic metastable vortex states in interacting vortex lines}
\author{Sergei Kozlov}
\affiliation{%
 Laboratoire de Physique et d'étude des Matériaux, ESPCI Paris, CNRS, PSL University, 75005 Paris, France
}

\author{Jérôme Lesueur}
\affiliation{%
 Laboratoire de Physique et d'étude des Matériaux, ESPCI Paris, CNRS, PSL University, 75005 Paris, France
}%

\author{Dimitri Roditchev}
\affiliation{%
 Laboratoire de Physique et d'étude des Matériaux, ESPCI Paris, CNRS, PSL University, 75005 Paris, France
}%

\author{Cheryl Feuillet-Palma}
\email{cheryl.palma@epsci.psl.eu}
\affiliation{%
 Laboratoire de Physique et d'étude des Matériaux, ESPCI Paris, CNRS, PSL University, 75005 Paris, France
}%

\date{\today}

\begin{abstract}
The electron transport in current-biased superconducting nano-bridges is determined by the motion of the quantum vortex confined in the internal disorder landscape.  Here we consider a simple case of a single or two neighbouring linear defects crossing a nano-bridge. The strong anharmonicity of the vortex motion along the defect leads, upon RF-excitation, to fractional Shapiro steps. In the case of two defects, the vortex motion becomes correlated, characterized by metastable states that can be locked to a resonant RF-drive. The lock-unlock process causes sudden voltage jumps and drops in the voltage-current characteristics observed in experiments. We analyze the parameters promoting these metastable dynamic states and discuss their potential applications in quantum devices.
\end{abstract}


\maketitle

\section{\label{sec:Introduction}Introduction}

Quantum vortices are famous topological objects – lines of 2$\pi$--phase singularities in the many-body wave function of coherent quantum condensates. In superconductors, where condensed particles are electrically charged Cooper pairs, the phase gradients generate vortex currents circulating around singularities, and producing a magnetic flux. The vortices strongly influence the characteristics of superconductors, limiting their critical currents and fields. In fact, externally applied currents and fields interact with the vortex, forcing it to move. In the vortex centers – cores – the superconductivity is suppressed, and the normal state is recovered. The vortex motion is therefore dissipative, often triggering the transition to the normal state of the entire system.

In their motion inside superconductors, vortices interact with various local and extended defects, as well as with other vortices and obstacles. 
The collection of defects along with other moving and pinned vortices form a potential landscape in which a given vortex evolves. This landscape is generally dynamic and intricate, comprising local minima and saddle points. Consequently, the formation of various metastable states can occur, with their characteristic energies and stability subject to perturbation by external magnetic fields, DC or AC currents.

In this work, we focus on the over-critical behaviour of current-biased superconducting nano-bridges (see Fig.\ref{fig:schema}a as an example) aiming to explain the spectacular results of recent experimental findings. In most of the cases, a nano-bridge behaves like a Josephson junction: when the critical current $I_c$ is reached, the bridge transits to the normal state. However, this transition is not always abrupt: the voltage $V(I>I_c)$ across the bridge increases progressively, often over several orders of magnitude, before the device reaches a fully normal state. In this progressive transition, the differential resistance $dV/dI(I)$ can exhibit drops, telegraphic "noise" behaviour, and even become negative \cite{zybtsev2006instabilities, buh2017phase, dobrovolskiy2017mobile}. When a microwave excitation is added, the nano-bridges display, similarly to Josephson junctions, the famous Shapiro steps in DC $V(I)$ characteristics; both integer and fractional plateaus are observed \cite{dinsmoreFractionalOrderShapiro2008, gubankov1976coherent, schneider1993nanobridges, rudenko1991stimulation, nawaz2013microwave}. To gain a microscopic insight into the physical origin of the observed phenomena, we provide Time-Dependent Ginzburg-Landau (TDGL) calculations in a superconducting nano-bridge in which only a few typical vortex pinning centers -- grain boundaries -- are present \cite{maggio1997critical}. We show that such a minimalistic disorder landscape is enough to explain several experimental results related to correlated vortex motion in disordered nano-bridges.

\section{\label{sec:Results}Results}

\subsection{\label{sec:Model}Model}

\begin{figure}
    \includegraphics{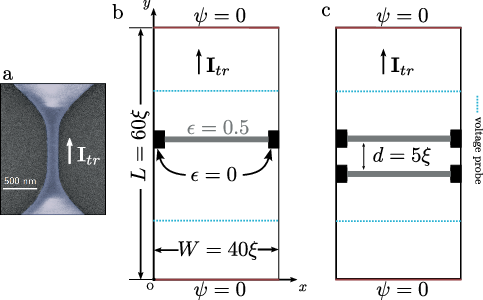}
    \caption{
    \label{fig:schema}
\textbf{Defects in superconducting nano-bridges.} \textbf{a} Scanning Electron Microscope image of a typical nano-bridge studied in experiments \cite{Amari}. \textbf{b} and \textbf{c} Two sample geometries studied theoretically, representing the central (narrowest) part of the real device. They contain one (\textbf{b}) or two (\textbf{c}) linear defects (grey regions). Edge defects situated at the ends of linear defects are indicated by black rectangles. The direction of the transport current is showed by arrows. The voltage is calculated between the two blue dashed lines. Further details are provided in the text.}
\end{figure}

The model system is a rectangular superconducting sample representing the central part of a typical nano-bridge, as shown in Fig. \ref{fig:schema}. The model bridge has a length of $L=60\xi$ and a width of $W=40\xi$, where $\xi$ represents the Ginzburg-Landau (GL) coherence length. 
Within the TDGL framework, the temporal and spatial evolution of the complex superconducting order parameter $\psi(t,\mathbf{r})$ can be expressed as \cite{sadovskyyStableLargescaleSolver2015} (see Methods):
\begin{equation}
    \begin{aligned}
    \label{eq:TDGL}
    &\partial_t \psi= \epsilon(\mathbf{r}) \psi-|\psi|^2 \psi+(\nabla-\imath \mathbf{A})^2 \psi \\
    &\varkappa^2 \nabla \times(\nabla \times \mathbf{A})=\mathbf{J}_{\mathrm{S}}+\mathbf{J}_{\mathrm{N}},
    \end{aligned}
\end{equation}
where $\mathbf{A}$ is the vector potential due to magnetic field,  $\mathbf{J}_S$ and $\mathbf{J}_N$ are superconducting and normal components of the electric current. The GL parameter $\varkappa=\lambda /\xi$ ($\lambda$ is the field penetration depth) is taken equal to $\varkappa=4$, meaning that the bridge is in the type-II regime (see Methods). 

The defects, as depicted in Fig.\ref{fig:schema}b,c, are introduced by spatially varying the parameter $\epsilon(\mathbf{r})$, which is associated with the local critical temperature $T_c(\mathbf{r})$ and the global sample temperature $T$:
\begin{equation}
    \label{eq:epsilon}
    \epsilon(\mathbf{r}) = \frac{T_c(\mathbf{r}) - T}{T} 
\end{equation}
The superconducting part of the bridge is described by $\epsilon$=1, while the defects are characterized by a locally reduced critical temperature $T_c(\mathbf{r})$, and are described by a lower $\epsilon (\mathbf{r})$. For instance, the linear (grey) defect of width 1$\times \xi$ crossing the bridge is characterized by $\epsilon = 0.5$. It represents an extended structural defect - a grain boundary crossing the real sample or an artificial weak-link (possible experimental realizations are discussed in Sec.\ref{sec:Discussion}). At a given temperature $T$, this defect is superconducting, but its local critical temperature is 3/4 of the critical temperature $T_c$ in the rest of the sample. The two point defects at the edges, presented by black rectangles $2\xi \times 5\xi$, are characterized by $\epsilon = 0$, that corresponds to a fully suppressed superconductivity. These edge defects appear at the ends of the grain boundaries as a result of a damage caused during the nano-bridge fabrication processes.

When simulating the Shapiro step experiments, the microwave illumination is added as an AC-current of amplitude $I_{_{AC}}$ and frequency $f_{_{AC}}$. The total transport current through the bridge is therefore: 
\begin{equation}
    \label{eq: currents}
I_{tr} = I_{_{DC}} + I_{_{AC}} \, sin(2\pi f_{_{AC}}t),
\end{equation}
The state of the bridge is determined by calculating the voltage $V$ between $y=0$ and $y=L$ boundaries for each value of transport current (see Methods). By averaging this voltage over the sample width and time one gets the DC-voltage $\langle V \rangle$ measured in experiments.

\subsection{\label{sec:One_line}Single linear defect}

As a starting point, we consider a single linear defect, as shown in Fig.\ref{fig:schema}b, that simulates a grain boundary crossing the bridge. Additionally, two point defects are introduced at the ends of the linear defect, representing suppressed superconductivity in the locations where the grain boundary reaches the sample edges. 

\begin{figure}
    \includegraphics{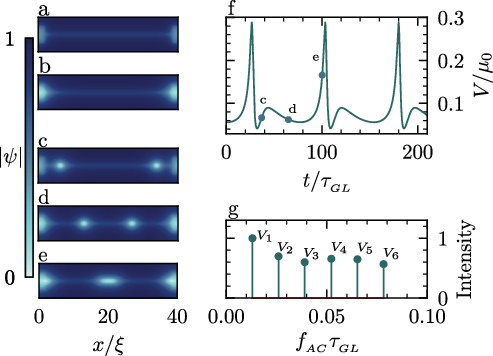}
    \caption{\label{fig:One-line-Vt} 
\textbf{Vortex dynamics in single linear defect with no AC current applied ($I_{_{AC}}=0$). a} Static map of the order parameter amplitude $|\psi(\mathbf{r})|$ at low transport currents $I_{_{DC}} \ll I_c$. \textbf{b} Static $|\psi(\mathbf{r})|$ map at $I_{_{DC}}\lesssim I_C$ indicates the presence of one vortex and one anti-vortex at the edge defects, ready to enter. \textbf{d-e} Snapshots of $|\psi(\mathbf{r})|$ at different moments of vortex propagation for a fixed $I_{_{DC}}$ = 0.10 $>$ $I_c$. \textbf{f} Periodic temporal evolution of the instantaneous voltage $V(t)$ at the same conditions. The dots c, d and e on the graph correspond to snapshots \textbf{c}, \textbf{d} and \textbf{e}. The period of $V(t)$ oscillations provides the fundamental frequency $f_1$ of the process. \textbf{g} Fourier spectrum of $V(t)$.}
\end{figure}

The results of calculations are presented in Fig.\ref{fig:One-line-Vt}. When the transport current $I_{tr}$ is well below a critical value $I_{c}$, the order parameter in the bridge is steady. It is depleted at the two local edge defects and at the linear defect, Fig.\ref{fig:One-line-Vt}a, following the imposed $\epsilon(\mathbf{r})$. At $I_{tr}\lesssim I_{c}$, there is already one vortex and one anti-vortex pinned at the two edge defects. The entire bridge remains in the superconducting state, with $V$=0, as expected. Figs.\ref{fig:One-line-Vt}c-e are snapshots of the temporal evolution of the order parameter amplitude when a constant $I_{tr}=0.10 > I_{c}$ is applied. Under this condition, one vortex and one anti-vortex simultaneously enter the bridge, Fig.\ref{fig:One-line-Vt}c. They accelerate towards each other under the action of the Lorenz force and experience mutual attraction, Fig.\ref{fig:One-line-Vt}d, and annihilate, Fig.\ref{fig:One-line-Vt}e. The process is periodic, with the period and details of the vortex-antivortex dynamics depending on the TDGL parameters. Moving vortices dissipate energy and generate an instantaneous voltage $V(t)$ proportional to the relative vortex velocity. Fig.\ref{fig:One-line-Vt}f illustrates the evolution of $V(t)$, with points c-e correspond to snapshots in Figs.\ref{fig:One-line-Vt}c-e. At the moment (c), the voltage rapidly rises as vortices accelerate due to their interaction with the edges and the transport current. In (d), $V$ crosses a local minimum as the vortex velocity drops in a region where interaction with the edge is already sufficiently small, and the transport current is reduced on the scale of $\sim \lambda$. A sharp increase in the vortex velocity due to the vortex-antivortex attraction just before annihilation produces a peak in $V(t)$ at moment (e). The Fourier spectrum of $V(t)$ is presented in Fig.\ref{fig:One-line-Vt}g. It contains the fundamental frequency of an amplitude $V_1$ and several harmonics with comparable amplitudes. Both $V(t)$ and spectrum indicate the strong anharmonicity of the vortex motion. It is important to note that the fundamental frequency is not fixed but grows with $I$ \cite{al2022driven} as the increasing Lorentz force pushes vortices to move and annihilate faster.

\begin{figure}
    \centering
    \includegraphics{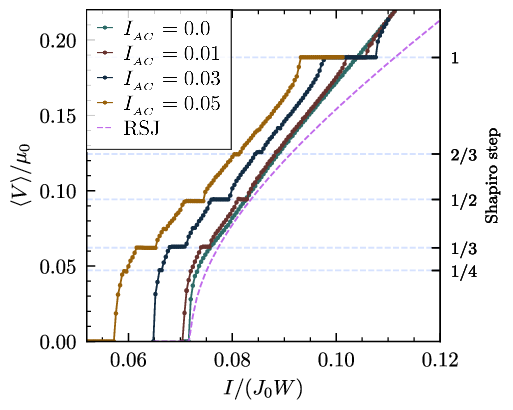}
    \caption{\label{fig:one_GB_IV}
\textbf{Transport properties of the nano-bridge with one linear defect.} Solid lines - normalized $\langle V \rangle (I_{_{DC}})$ characteristics calculated for different values of the AC component $I_{_{AC}}$ of the total transport current. The right vertical axis displays the numbers of Shapiro plateaus. 
Dashed line - $V(I)$ characteristic of a SNS Josephson junction calculated within the RSJ model \cite{stewart1968current, mccumber1968effect}.  To fit the curve into the plot window, its vertical scale was divided by a factor of 10.
    }
\end{figure}

By repeating the calculations for different $I_{_{DC}}$ and time-averaging $\langle V(t) \rangle$, we retrieve $\langle V \rangle (I_{_{DC}})$ dependencies measured in experiments. The dark green curve in Fig.\ref{fig:one_GB_IV} is the result of these calculations for the range of $I_{_{DC}}$ around the transition from the non-dissipative to the dissipative state. The plots are presented in reduced coordinates $\langle V \rangle/\mu_0$ vs $I_{_{DC}}/(J_0W)$ (see Methods). The shape of the curve resembles the $V(I)$ characteristics of an ordinary Superconductor -- Normal metal -- Superconductor (SNS) Josephson junction. The latter, represented by the dashed line in Fig.\ref{fig:one_GB_IV}, was calculated using the Restively Shunted Junction (RSJ) model. Both curves exhibit a non-dissipative branch at low currents, a rise at some critical current, and a smooth increase at higher currents. However, the resemblance is limited. First, in the present case, the SNS junction does not exist, and the intrinsic critical (depairing) current of the bridge $\sim J_0W$ is much higher than the calculated value $I_c\simeq0.072J_0W$. Second, in SNS junctions, the $\langle V \rangle (I_{_{DC}})$ curve asymptotically approaches the normal branch $\langle V \rangle=R_N I_{_{DC}}$ as $I_{_{DC}}$ increases, while the bridge "resistance" $\langle V \rangle /I{_{DC}}\times (J_0W/\mu_0)$ remains much lower than its normal state resistance $R_N$ (this is why the vertical scale of the RSJ curve was reduced to fit it in the plot window). Third, the RSJ model fails in reproducing an almost linear rise of $\langle V \rangle (I_{_{DC}})$ for $I > I_c$. The three deviations originate from the fact that, contrary to SNS junctions in which the voltage appears as a result of the suppression of the proximity-induced superconducting correlations in the N-part, in the nano-bridges it is due to individual vortex motion inside a still superconducting device. This difference is essential, leading to unique transport properties that we focus on in this work.

The SNS-like behaviour of the bridge is further evidenced by simulating its response to microwave illumination. When an AC-current is added, the oscillating voltage $V(t)$ can be locked to the frequency $f_{_{AC}}$ of this external drive, resulting in plateaus of constant voltage on $\langle V \rangle (I_{_{DC}})$ curve, as shown in Fig.\ref{fig:one_GB_IV}. This effect resembles the well-known Shapiro steps observed in ordinary Josephson junctions under microwave illumination, where the $n^{th}$ voltage plateau is defined by the locking condition $f_1 = n \cdot f_{_{AC}}$ (where $n$ is an integer) and the second Josephson relation $\langle V(t) \rangle = hf_{1}/2e$, where $f_1$ is the fundamental (Josephson) frequency.

In addition to the integer Shapiro plateaus, the fractional ones are also revealed. These plateaus appear in Fig.\ref{fig:one_GB_IV} at voltages satisfying the condition $\frac{n}{k} \cdot hf_{_{AC}} = 2e\langle V(t) \rangle$, where $n$ and $k$ are integers. The presence of fractional plateaus is directly linked to a high anharmonicity of $V(t)$ oscillations in Fig.\ref{fig:One-line-Vt}f. The Fourier spectrum of $V(t)$, presented in Fig.\ref{fig:One-line-Vt}g, is indeed characterized by high amplitudes $V_k$ of $k^{th}$ voltage harmonics (at a frequency $f_k$) which are comparable to the amplitude $V_1$ at its fundamental frequency $f_1$. This enables an efficient locking of these harmonics to the AC-drive when $f_k=kf_1= n f_{_{AC}}$.

\begin{figure}[h!]
    \centering
    \includegraphics{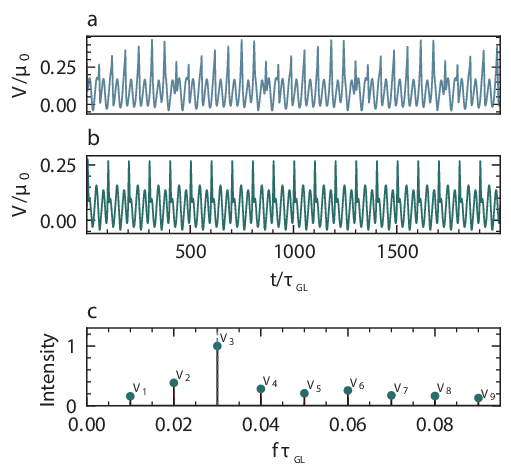}
    \caption{\label{fig:one_GB_spectrum}
   \textbf{Evolution of the instantaneous voltage $V(t)$ for the Shapiro plateau $1/3$ of Fig.3. a} $V(t)$ at constant $I_{_{DC}}$ and $I_{_{AC}}$ for detuned frequency $f_{_{AC}}=0.625 \, f_3$. The low-frequency envelope due to the beat effect is visible. \textbf{b} $V(t)$ at the resonance $f_{_{AC}}=f_3$.  \textbf{c} Frequency spectrum of $V(t)$ in the case (\textbf{b}). }
\end{figure}

In Fig.\ref{fig:one_GB_spectrum}, the instantaneous voltage $V(t)$ and its spectrum for the Shapiro plateau $1/3$ are shown. When the frequency $f_{_{AC}}$ is slightly detuned  from $f_k/n$, a low-frequency envelope of a beat frequency $|f_{_{AC}} - f_k/n|$ is observed. This effect allows for the detection of higher harmonics experimentally, even when their magnitude is small and imperceptible in $\langle V \rangle (I_{_{DC}})$ curves in the Shapiro step experiment. Consequently, by tunning the amplitude and frequency of the AC excitation, it becomes possible to induce fractional Shapiro step when $f_{_{AC}}$ is not only a multiple of $f_1$ but a multiple of higher $f_k$ harmonics. All these features reveal a rich spectral characteristic of the considered system. It should be mentioned that when the amplitude $I_{_{AC}}$ becomes comparable to $I_{_{DC}}$, the AC-excitation cannot be considered as a perturbation anymore. Instead, one should think of a complex dynamical system whose spectrum (amplitudes $V_k$ and frequencies $f_k$) depends on both components of $I_{tr}$.

Concluding this section, it is important to recall pioneering works \cite{aslamazov1975josephson, likharev1975steady} that predicted similarities between the transport properties of Josephson junctions and those of nano-bridges crossed by vortices (or phase-slips). These predictions were later confirmed in several experiments, where both integer \cite{gubankov1976coherent, sivakovJosephsonBehaviorPhaseSlip2003, schneider1993nanobridges, nawaz2013microwave} and fractional \cite{dinsmoreFractionalOrderShapiro2008} Shapiro steps were observed. This analogy was also explored, both experimentally and theoretically, in the case of vortices jumping between pinning sites \cite{fioryQuantumInterferenceEffects1971, martinoliStaticDynamicInteraction1978, van1999shapiro, reichhardt2000phase,  al2022driven}.

\subsection{\label{sec:Two_line}Two neighbouring linear defects}

In experimentally studied nano-bridges, the disorder is rarely represented by only one grain boundary. Most non-epitaxial superconducting films exhibit granularity on a scale of 20-200 nm, which can be significantly shorter than the nano-bridge width $W$. For instance, nano-meanders studied in \cite{amari2017high} were elaborated out of thin $\mathrm{YBa}_2\mathrm{Cu}_3\mathrm{O}_{7-\delta}$ films. They possess a specific morphology \cite{mannhart1996generation} and form a network of grain boundaries. Statistically, several such boundaries can cross the bridge. The vortex motion in these networks is much more complex than in a single grain boundary studied above. While the vortex cores are confined within the grain boundaries \cite{maggio1997critical}, the vortex currents extend far beyond; they circulate on a scale of $\lambda$ or, in ultrathin films, on an even larger scale of the Pearl penetration depth \cite{pearlCURRENTDISTRIBUTIONSUPERCONDUCTING2004a}. This results in a mutual interaction between vortices present in different grain boundaries, affecting their collective motion. As a step towards accounting for this complexity, we consider now two linear defects (grain boundaries) characterized by $\epsilon$=0.5. The defects are separated by a distance $l=5\xi \sim \lambda$, Fig.\ref{fig:schema}c, thus inducing an interline vortex-vortex interaction.

\begin{figure}[h!]
    \centering
    \includegraphics{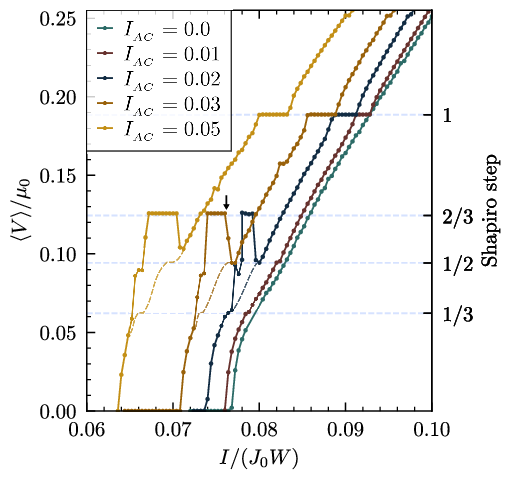}
    \caption{\label{fig:two_GB_IVs}
    \textbf{$\langle V \rangle (I_{_{DC}})$ characteristics for different values of $I_{_{AC}}$ in the case of two identical linear defects (displayed in fig.\ref{fig:schema}c).} The right vertical axis displays the numbers of Shapiro steps. Dashed lines are used as eye-guides (see in the text).}
\end{figure}

\begin{figure}
    \centering
    \includegraphics{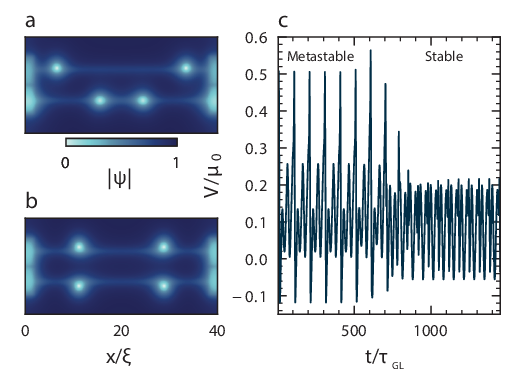}
    \caption{\label{fig:Vt_transition}
\textbf{Vortex dynamics in the case of two identical linear defects of fig.\ref{fig:schema}c. a} and \textbf{b} Snapshots of the order parameter amplitude in the stable (\textbf{a}) and metastable (\textbf{b}) states near the transition (see in the text). \textbf{c} Evolution of $V(t)$ at the transition from the metastable to the stable state at $I_{_{AC}}=0.03$. The initial DC-current $I_{_{DC}}=0.076$ switches to $I_{_{DC}}=0.077$ at the moment $t$=500. }
\end{figure}

The calculated $\langle V \rangle (I_{_{DC}})$ characteristics in the case of two identical linear defects is presented as a dark green solid line in Fig.\ref{fig:two_GB_IVs}. The shape of this curve is almost identical to that obtained in the case of a single linear defect. As in the previous case, at DC-currents just above the critical one $I_{_{DC}} \gtrsim I_{c}$, one vortex-antivortex pair enters the nano-bridge and moves along one of the two linear defects, thus generating a non-zero voltage. The only difference with the single defect case is that after vortex-antivortex annihilation in one line, a new vortex-antivortex pair enters the other line, and the process repeats. Further increase of the DC-current leads to an acceleration of vortices and, consequently, to an increase of voltage $\langle V \rangle$. At a high enough DC-current, the system enters a new state in which the second vortex-antivortex pair enters into the second line before the first pair annihilates in the first one. This moment is witnessed by a slight inflexion of $\langle V \rangle (I_{_{DC}})$ curve at $I_{_{DC}} \simeq$0.084. In this state, there are two vortex-antivortex pairs in the nano-bridge at the same time. Due to the mutual repulsion of vortices of the same sign, they try to position themselves as far from each other as possible, while remaining inside linear defects. This leads to a lateral $x$-shift of the vortex positions in neighbouring lines, as shown in Fig.\ref{fig:Vt_transition}a. This dynamic vortex pattern is reminiscent of the static Abrikosov vortex lattice. As time advances, the vortex-antivortex pair in the bottom line annihilates, the one in the top line advances towards the center, and a new one enters the bottom line.

By adding a low AC-current, one gets the Shapiro plateaus that also look very similar to the single defect case (brown line in Fig.\ref{fig:two_GB_IVs}). Though, at higher AC-currents new features appear. These are large 2/3 Shapiro plateaus on $\langle V \rangle (I_{_{DC}})$ curves with a rapid voltage raise on their left side and a voltage drop on their right side (hand-added smooth dashed lines help to appreciate the amplitude of the effect). Unlike other plateaus, the width of the 2/3 plateau rapidly grows with the AC-current (compare the curves at $I_{_{AC}}= $0.02, 0.03, and 0.05). 

To understand the origin of this phenomenon, let us consider the dynamics of the system close to the voltage drops. In the specific case of $I_{_{AC}}=0.03$, this occurs at  $I_{_{DC}}^{drop} \simeq$ 0.076, as indicated by the arrow on the $\langle V \rangle (I_{_{DC}})$ curve in Fig.\ref{fig:two_GB_IVs}. 
The calculations show that just above $I_{_{DC}}^{drop}$, the vortex-antivortex motion in the two defects is sequential, as presented in Fig.\ref{fig:Vt_transition}a, while just below $I_{_{DC}}^{drop}$ (that is on the plateau) it is synchronous: Vortex-antivortex pairs enter the defects simultaneously, move in parallel to each other (see the snapshot Fig.\ref{fig:Vt_transition}b), and annihilate at the same time. This leads to high peak-to-peak voltage spikes in $V(t)$, as those visible on the left side of Fig.\ref{fig:Vt_transition}c.

The synchronous configuration is not stable itself. Indeed, when a vortex in one line is located under a vortex in the other, the projection of a vortex-vortex repulsion force on a $x$-axis is zero, and any $x$-shift of their position gives rise to the $x$-axis component of vortex-vortex repulsion which drives the system out of this unstable balance towards a more stable 
checkerboard configuration, Fig.\ref{fig:Vt_transition}a. Thus, the metastable configuration of Fig.\ref{fig:Vt_transition}b is stabilized by the external AC-drive that works as a periodic force; if the amplitude of this force (proportional to $I_{_{AC}}$) is sufficient, the configuration is stabilized, in some range of external parameters, giving rise to a plateau on $\langle V \rangle (I_{_{DC}})$ curve.
When the DC-current is slightly increased above $I_{_{DC}}^{drop}$, the Lorenz force  increases, and the system jumps down to the stable configuration of Fig.\ref{fig:Vt_transition}a. The corresponding evolution of $V(t)$ is presented in Fig.\ref{fig:Vt_transition}c.
One can observe that after a few periods of high peak-to-peak voltage oscillations, the system transits to oscillations with a nearly twice lower peak-to-peak voltage 
(compare left and right parts of Fig.\ref{fig:Vt_transition}c). This change is due to the fact that in the metastable state, the vortex-antivortex annihilation takes place simultaneously in the two lines, while in the stable configuration the process is sequential. The system is no more locked to the 2/3 Shapiro step in the stable configuration. A movie illustrating the oscillatory dynamics of this transition is provided in the Supplementary Material.

The motion of vortices in the two close linear defects can be seen as a system of two coupled identical anharmonic oscillators. In this representation, the two oscillation patterns of Fig.\ref{fig:Vt_transition} can be seen as two modes, one of which is low in energy ($E_0$) and therefore stable, while the other, at higher energy $E_1$, is metastable. Each of these modes depends on DC current $I_{_{DC}}$, and the evolution of the lowest mode corresponds to $\langle V \rangle (I_{_{DC}})$ curve at $I_{_{AC}} = 0$. Another mode can only be achieved with an external excitation, in a certain range of pumping powers and frequencies.

\begin{figure}
    \centering
    \includegraphics{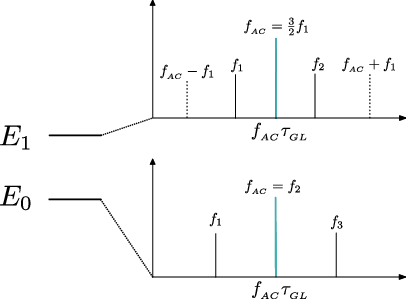}
\caption{\label{fig:diagramm}
        \textbf{Schematic energy diagram of the states considered in Fig.6}. Stable ($E_0$) and metastable ($E_1$) states are presented along with their frequency spectra.
    }
\end{figure}

The calculated Fourier spectra of $V(t)$ in the states $E_0$ and $E_1$ are presented in Fig.\ref{fig:diagramm}. In the metastable configuration $E_1$, the AC-drive locks to the third harmonic of the system as $2f_{_{AC}}=3f_{1}$. The Josephson frequency is $f_{1}=(2/3) f_{_{AC}}$ and, consequently, the DC-voltage measured in the experiment is  $\langle V(t) \rangle = (2/3) h f_{_{AC}}/2e$. This voltage remains constant as long as the system is locked to the drive, resulting in the unusual 2/3 Shapiro plateau in Fig.\ref{fig:two_GB_IVs}. Immediately after the drop, the drive locks to the second harmonic as $f_{_{AC}}=2f_{1}$, that is $f_{1}=(1/2) f_{_{AC}}$, resulting in a lower DC-voltage $\langle V(t) \rangle = (1/2) h f_{_{AC}}/2e$. In principle, it could be the usual 1/2 Shapiro plateau, due to anharmonicity. Though, when the AC-current increases and the width of the unusual 2/3 plateau rapidly grows, the plateau 1/2 shrinks and disappears (compare the curves at $I_{_{AC}}=$ 0.02, 0.03 and 0.05 in Fig.\ref{fig:two_GB_IVs}). Note that as $I_{_{DC}}$ is further increased above $I_{_{DC}}^{drop}$, the Lorentz force rises pushing vortices to move faster, the corresponding frequencies grow, and the lock to the fixed frequency of the AC-drive is lost. This roller-coaster ride between different metastable, stable locked and unlocked states is reflected in voltage spectra and as a consequence in a $\langle V \rangle (I_{_{DC}})$ curve.

\begin{figure}[h!]
    \centering
    \includegraphics{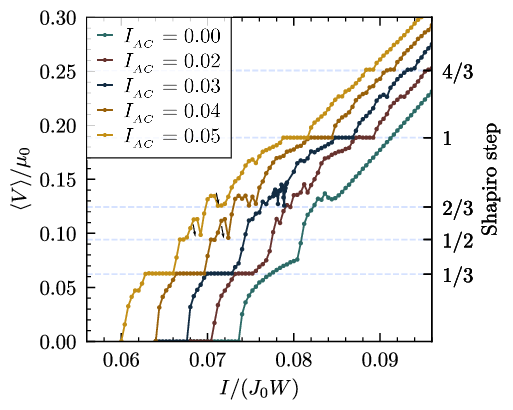}
    \caption{\label{fig:mismatch}
    \textbf{$\langle V \rangle (I_{_{DC}})$ characteristics for different $I_{_{AC}}$ in the case of two different linear defects, $\epsilon=0.5$ and $\epsilon=0.42$.} The right vertical axis displays the numbers of corresponding Shapiro steps.}
\end{figure}

Till now we have considered a very idealistic case where the two coupled linear defects were identical. This situation could be realized in artificial stacks of SNS junctions \cite{berdiyorov2013synchronized}, periodic pinning arrays \cite{al2005chaotic, van1999shapiro} but not in nano-bridges made of films in which the intrinsic pining landscape is aperiodic and the inter-grain coupling varies from one grain boundary to the other. To account for this diversity, we also studied asymmetric linear defects. In Fig.\ref{fig:mismatch}, we show the $\langle V \rangle (I_{_{DC}})$ characteristics for the case of two linear defects located as in Fig.\ref{fig:schema}c, but characterized by different $\epsilon$ parameters: $\epsilon=0.5$ and $\epsilon=0.42$. The curves differ significantly from the previous case, even without AC-excitation (green curve). The critical current is lower, and for 0.074 $<I_{_{DC}} <$ 0.0805, the voltage appears exclusively due to the vortex motion in the $\epsilon=0.42$ line; the $\epsilon=0.5$ line contains no vortices. Above $I_{_{DC}} \simeq$ 0.0805, the vortices start to penetrate the second line as well, and at high enough currents, $I_{_{DC}} \gtrsim$ 0.085, their motion becomes mutually synchronized, similarly to the previous case displayed in Fig.\ref{fig:Vt_transition}a. Remarkably, in the intermediate current region, 0.0805 $< I_{_{DC}} <$ 0.085, the two anharmonic oscillators have very different spectral fingerprints and, as a result, there is no clear synchronization of the vortex motion in the two lines; in this region, the $\langle V \rangle (I_{_{DC}})$ characteristics demonstrates a bump with several local maxima and minima. When AC-excitation is added, integer and fractional Shapiro steps are observed, the latter stemming from the anharmonic nature of vortex motion. The transitions to/from metastable modes are also observed, although their number is larger, their shape more complex and intricate than in the case of identical defects. Clearly, the vortex dynamics in the presence of asymmetric defects leads to a greater variety of collective motion modes.

\section{\label{sec:Discussion} Discussion}

The evolution of Shapiro features in Figs.\ref{fig:two_GB_IVs},\ref{fig:mismatch} with increasing $I_{_{AC}}$ is not trivial. At low AC-excitation, $I_{_{AC}} \ll I_{_{DC}}$, conventional Shapiro plateaus are narrow, and no signatures of metastable states are seen. In this regime, the AC-component acts as a probe that locks, at a fixed $f_{_{AC}}$, onto the spectrum of the vortex motion, solely determined by the main driving (Lorentz) force $\sim I_{_{DC}}$. As $I_{_{DC}}$ increases, the vortices move faster, $f_1$ and $f_k$ increase. At some $I_{_{DC}}$, a given $f_k$ gets close enough to $nf_{_{AC}}$, and the motion locks to $f_{_{AC}}$; $f_1$ remains fixed in some range of $I_{_{DC}}$. As $I_{_{DC}}$ further increases, the locking effect is lost. This results in a series of integer and fractional Shapiro plateaus visible on $\langle V \rangle (I_{_{DC}})$ curve at $I_{_{AC}}$=0.02.

When $I_{_{AC}}$ is increased and becomes comparable with $I_{_{DC}}$, two phenomena appear. The first one is the well-known enlargement of Shapiro plateaus, due to a stronger locking effect at higher AC-currents. The second one is related to the perturbation of the vortex motion spectrum by the oscillatory force $\sim I_{_{AC}}$, whose amplitude becomes comparable to the Lorentz force due to DC-current. The combined action of $I_{_{DC}}$ and $I_{_{AC}}$ enables the existence of metastable states. They can be locked to $f_{_{AC}}$, resulting in jump-plateau-drop features as observed in Fig.\ref{fig:two_GB_IVs}. The same phenomenon takes place in Fig.\ref{fig:mismatch}, where many more voltage bumps and drops are observed (some jumps to metastable states are indicated by black arrows) as compared to Fig.\ref{fig:two_GB_IVs}. The lift of degeneracy, resulting in a more rich and complex metastable state spectrum, is certainly behind these differences. Finally, the DC-current range where the feature appears rapidly extends with increasing $I_{_{AC}}$. 

In the limit of a dense, on the scale of $W$, network of defects, one would expect a huge number of apparently chaotically arranged voltage jump-bump-drops to appear on $\langle V \rangle (I_{_{DC}})$ curves, reflecting a vast number of accessed vortex motion modes and the complexity of the related spectra. The term "chaotic" is justified here due to a high sensitivity of the accessed metastables configurations to external parameters such as $I_{_{DC}}$, $I_{_{AC}}$, $f_{_{AC}}$, the disorder landscape, etc. Indeed, after unlocking from one metastable state, the system can jump down to a more stable configuration or lock up to another metastable state, from the available set. As a result, the position and shape of bumps-drops on $\langle V \rangle (I_{_{DC}})$ curves would appear arbitrary (see Fig.\ref{fig:mismatch}), while they are deterministic.

The revealed voltage drops correspond to a negative dynamic resistance $dV/dI(I_{_{DC}})$. The latter has been experimentally observed in periodic pinning arrays subject to a specific external magnetic field \cite{gutierrez2009transition, misko2006nonuniform, reichhardt1997dynamic}, where a complex collective dynamics of vortices led to multiple phase transitions in their collective motion, with no need for additional AC-drive, resulting in various features in the $V(I)$ characteristics \cite{gutierrez2009transition, misko2006nonuniform, reichhardt1997dynamic}. Another system is a perforated Nb film put in an external magnetic field, where the negative dynamic resistance can appear due to the Ratchet effect under AC-drive \cite{dobrovolskiy2017mobile}. More recently, both Shapiro steps and negative dynamic resistance were observed in a MoN strips with an artificial cut  \cite{ustavschikov2022negative}. The authors attributed the negative dynamic resistance to the chaotic aperiodic vortex motion at high AC-excitation amplitude.

The ability to use AC-excitation both as a pump and as a probe opens up interesting possibilities for realization, spectroscopy and control of metastable states in superconducting weak-links. The obtained results demonstrate the potential for designing artificial disorder landscapes to achieve desired responses to AC-amplitude and/or frequency. The general nature of weak-links suggests that there would be multiple ways of experimental realization of these functionalities. One of straightforward routes is to engineer superconducting films with a controlled disorder by using Focused Ion Beam approaches \cite{dobrovolskiy2017mobile} or to deposit superconducting materials onto faceted structures \cite{soroka2007guiding}. By carefully designing the spatial distribution of defects or grain boundaries, one could tailor the response of the weak-links to both DC- and AC-excitation. Another avenue is to overlap superconducting weak-links by ferromagnetic strips that can locally suppress the superconducting order parameter due to the inverse proximity effect \cite{carapella2016current, carapella2016mesoscopic, dobrovolskiy2020upper, jaque2002anisotropic, yuzhelevski1999artificial}. This can introduce additional complexity in the vortex dynamics and lead to novel effects under microwave excitation. 
\section{Conclusion}

In this work, we numerically studied transport properties of current-carrying superconducting nano-bridges subject to microwave illumination. The granularity of experimentally measured devices was accounted for by introducing one or two linear defects (simulating grain boundaries) which were directed perpendicularly to the applied current. We revealed a rich and complex dynamics of the vortex motion along these defects. Its strong anharmonicity enabled us to lock the spectrum of the system to an external periodic drive, and to obtain both integer and fractional Shapiro plateaus in DC voltage-current characteristics. In the case of two close linear defects, the inter-vortex coupling leads to the appearance of collective modes of correlated motion, with multiple stable and metastable states. These transitions are revealed in the current-voltage characteristic as regions of negative differential resistance $dV/dI(I_{_{DC}})$. By playing with the external drive amplitude and frequency it becomes possible to pump the system to higher-resistance metastable modes and stabilize it there, in a finite range of DC transport currents. A step out of this range leads to a relaxation to a lower-resistance modes. The ability to control and stabilize different modes of the vortex motion opens up new possibilities for designing superconducting devices with tunable transport properties and novel functionalities.

\section{\label{sec:METHODS}METHODS}

Within the TDGL framework, the temporal and spatial evolution of the complex superconducting order parameter $\psi(t,\mathbf{r})$ can be expressed as \cite{sadovskyyStableLargescaleSolver2015}:
\begin{equation}
    \begin{aligned}
    &u\left(\partial_t+\imath \mu\right) \psi= \epsilon(\mathbf{r}) \psi-|\psi|^2 \psi+(\nabla-\imath \mathbf{A})^2 \psi \\
    &\varkappa^2 \nabla \times(\nabla \times \mathbf{A})=\mathbf{J}_{\mathrm{S}}+\mathbf{J}_{\mathrm{N}},
    \end{aligned}
\end{equation}
where $\psi$ is in units of $\psi_0 = \sqrt{\frac{|a|}{b}}$, with $a$ and $b$ being phenomenological parameters of the GL theory. The parameter $u$=1 is taken since we focus only on vortex motion but not on its nucleation dynamics. The coordinates $\mathbf{r}=(x,y)$ are in units of $\xi$. The scalar potential $\mu$ is measured in units of $\mu_{0} = \frac{\hbar}{2e\tau_{{GL}}}$, where $\tau_{_{GL}} = \frac{4\pi\sigma\lambda^{2}}{c^2}$ denotes the GL relaxation time, and $\lambda$ is the London penetration depth. The parameter $\sigma$ corresponds to the normal state conductivity of the material. The variable $t$ is measured in units of $\tau_{_{GL}}$, while the vector potential $\mathbf{A}$ is in units of $H_{c_2}\xi$, with $H_{c_2} = \frac{\hbar c}{2e\xi^2}$ representing the upper critical field. The parameter $\epsilon(\mathbf{r})$ is associated with the local critical temperature $T_c(\mathbf{r})$ through Eq.\eqref{eq:epsilon}; it enables spatially modulating the strength of the order parameter.

In the second GL equation, the total current $\mathbf{J}$ has superconducting ($\mathbf{J}_S$) and a normal ($\mathbf{J}_N$) components; it can be expressed in units of $J_{0} = \frac{c \Phi_0}{8 \pi^2 \lambda^2 \xi}$ as:
\begin{equation}
\mathbf{J}=\mathbf{J}_{\mathrm{S}}+\mathbf{J}_{\mathrm{N}}=\operatorname{Im}\left[\psi^*(\nabla-\imath \mathbf{A}) \psi\right]-\left(\nabla \mu+\partial_t \mathbf{A}\right)
\end{equation}

As TDGL equations are invariant under a gauge transformation, we use the zero scalar potential $\mu = 0$ gauge to eliminate the scalar potential from both equations. For simplicity, we set the order parameter equal to zero $\psi = 0$ on boundaries $y=0, L$. To apply external transport current $I_{tr}$, we use boundary conditions for the vector potential on boundaries $x=0, W$ as $\nabla \times \mathbf{A}$ = \textbf{\textit{H$_I$}}, where $H_{I}=2\pi I_{tr}/c$ represents the magnetic field induced by the transport current. On the other boundaries, we set $\nabla \times \mathbf{A} = \mathbf{0}$. Additionally, we impose the superconductor-vacuum boundary condition $\mathbf{n} \cdot (\nabla-\imath \mathbf{A}) \psi = 0$ on boundaries $x=0, W$, where $\mathbf{n}$ is the normal vector to the boundaries.

The state of the bridge is determined by calculating the voltage $V$ between $y=0$ and $y=L$ boundaries for each value of transport current. In the chosen gauge, the electric field is written as $\mathbf{E} = -\partial_t \mathbf{A}$. The corresponding instantaneous voltage drop $V_{y_1,y_2}$ between two arbitrary points $y_1, y_2$ in $y$-direction can be calculated as 
\begin{equation}
    V_{y_1,y_2}(x,t)=-\int_{y_1}^{y_2} E_y(x,y,t) \,dy = \int_{y_1}^{y_2} \partial_t A_y(x,y,t) \,dy.
\end{equation}
By averaging this voltage over the sample width and time we get the DC-voltage $\langle V \rangle$ measured in experiments. To avoid the voltage drops at $y=0, L$ boundaries, we calculate the voltage inside the bridge where the order parameter is fully restored ($\psi = 1$), as indicated by the blue dashed lines in Fig.\ref{fig:schema}. When simulating the Shapiro step experiments, we consider the microwave illumination as an additional time-dependent transport current of amplitude $I_{_{AC}}$ and frequency $f_{_{AC}}$. The total transport current through the bridge is given by Eq.\eqref{eq: currents}.

In the model, all non-equilibrium quasiparticle processes are omitted, and for all considered frequencies, the microwave illumination acts on vortices only as an additional periodic Lorenz force.

The system of Eqs.\eqref{eq:TDGL}, with the above-described boundary conditions, was solved using the commonly used link-variable method \cite{machidaDirectSimulationTimedependent1993, groppNumericalSimulationVortex1996, winieckiFastSemiImplicitFinite2002, sadovskyyStableLargescaleSolver2015} on the finite-difference grid. Spatial derivatives were approximated using the central difference method, and for time integration, the forward Euler method was employed \cite{langtangen2017finite}. In all calculations of Shapiro steps, we set $f_{_{AC}}\tau_{_{GL}}=0.03$.

\begin{acknowledgments}
We thank A. Gurevich for fruitful discussions and L. di Medici for sharing his computational power. 
This project has received funding from the European Union’s Horizon 2020 research and innovation programme under the Marie Skłodowska-Curie grant agreement No 754387. This work has been supported by the ANR JCJC (HECTOR ANR-21-CE47-0002-01), by Thales through a Co-fund PhD fellowship and was granted access to the HPC resources of MesoPSL financed by the Region Ile de France the project Equip@Meso (reference ANR-10-EQPX-29-01) of the programme Investissements d’Avenir supervised by the Agence Nationale pour la Recherche.
\end{acknowledgments}

\section{Author contributions}
S.K, C.F.P. and D.R. proposed the idea, S.K. conducted all the numerical simulations in the Time Dependant Ginzburg-Landau framework under the guidance of C.F.P, D.R. All the authors discussed the numerical results and participated in writing of the manuscript.

\end{document}